# EXCITATION OF A FUNNEL-SHAPE OPTICAL NEAR FIELD BY THE LAGUARRE-GAUSSIAN DOUGHNUT BEAM

S. M. Iftiquar[A], Haruhiko Ito[A,B], Akifumi Takamizawa[B] and Motoichi Ohtsu[A,B]

[A] ERATO Localized Photon Project, Japan Science and Technology Corporation, 687-1-17/4F Tsuruma, Tokyo 194-0004, Japan. E-mail: iftiquar@ohtsu.jst.go.jp

[B] Interdisciplinary Graduate School of Science and Engineering, Tokyo Institute of Technology, 4259 Nagatsuta, Kanagawa 226-8502, Japan

For the precise control of atoms beyond the diffraction limit, we consider the use of optical near fields localized in a nanometer region [1]. In this case, high-density cold atoms are required for effective interaction with the nanometric optical near fields. To this end, we are developing an atom funnel, which makes a cold atomic beam from laser-cooled atoms [2]. The atom funnel is composed of an inverse triangular-pyramidal hollow prism with a 200-microns exit hole. The optical near field produced on the inner-wall surface by a doughnut-shaped light beam reflects and cools atoms falling from a magneto-optical trap and collects them at the bottom under a blue-detuning condition. The doughnut beam with a power of more than 500 mW and a diameter of more than 4 mm are required for efficient funneling of Rb atoms. In addition, the dark centre of 200 microns is needed for avoidance of the spontaneous-emission heating [3]. In this paper, we report the generation of the optical near field using the Laguarre-Gaussian beam with a doughnut mode [4]. Figure 1 shows a cross-sectional image of a doughnut beam with a power of 0.5 W produced from a Gaussian one by means of interference [4] of spherical waves. The beam diameter is 5 mm, while the hollow diameter is about 250 microns. The method of Gouy phase difference used here has several advantages, compared to the conventional ones such as holography. First, the conversion efficiency is up to 50 %. Second, the diameter of the dark centre is variable and can be made below 100 microns. Thanks to these advantages, we can make the special doughnut beam with a large bright ring and a small dark centre. Moreover, it is possible to keep the hollow region over a long distance of more than 200 cm. Since the doughnut beam has an angular moment to the propagation direction, the atoms see the radiation pressure toward the same direction. Consequently, the atoms coming out of the exit hole can be confined by the transverse dipole force and guided by the longitudinal spontaneous force through the doughnut beam to an arbitrary point. Figure 2 shows the intensity profile of the optical near field (evanescent field) produced on a surface by the doughnut beam shown in Fig. 1. The peak intensity is estimated to be more than 5 W/cm$^2$. Assuming this optical near field, we can show that laser-cooled Rb atom of more than 50 % are converted to an atomic beam with a mean kinetic energy of about 100 microkelvin in terms of temperature. In this case, the atomic flux is estimated to be $10^{13}$ atom/s cm$^2$, which satisfies our conditions [3].

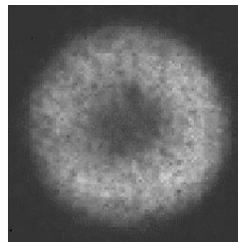

**Figure 1**
CCD image of a doughnut beam converted from a Gaussuan beam by the interference method.

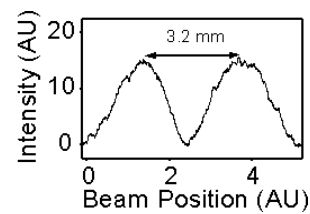

**Figure 2**
Cross-sectional intensity profile of an optical near field produced by the doughnut beam